# Trends in Energy Estimates for Computing in AI/Machine Learning Accelerators, Supercomputers, and Compute-Intensive Applications*

Sadasivan Shankar[1], Albert Reuther[2]

[1]SLAC National Laboratory, Menlo Park, CA; Materials Science and Engineering, Stanford University, CA, USA
{sshankar@slac.stanford.edu; sadas.shankar@stanford.edu}

[2]MIT Lincoln Laboratory Supercomputing Center (LLSC), Lexington, MA, USA
{reuther@LL.mit.edu}

*Abstract*—We examine the computational energy requirements of different systems driven by the geometrical scaling law (known as Moore's law or Dennard Scaling for geometry) and increasing use of Artificial Intelligence/ Machine Learning (AI/ML) over the last decade. With more scientific and technology applications based on data-driven discovery, machine learning methods, especially deep neural networks, have become widely used. In order to enable such applications, both hardware accelerators and advanced AI/ML methods have led to the introduction of new architectures, system designs, algorithms, and software. Our analysis of energy trends indicates three important observations: 1) Energy efficiency due to geometrical scaling is slowing down; 2) The energy efficiency at the bit-level does not translate into efficiency at the instruction-level, or at the system-level for a variety of systems, especially for large-scale AI/ML accelerators or supercomputers; 3) At the application level, general-purpose AI/ML methods can be computationally energy intensive, off-setting the gains in energy from geometrical scaling and special purpose accelerators. Further, our analysis provides specific pointers for integrating energy efficiency with performance analysis for enabling high-performance and sustainable computing in the future.

*Keywords—Moore's law, Energy Efficiency in computing, Energy per Instruction, Energy per Bit, Instructions per Second, Bit Utilization, Specialized Architectures, Energy for Machine Learning Application, Co-design, Energy as a design attribute*

## I. Introduction

As Artificial Intelligence and Machine Learning (AI/ML) methods evolve with advanced techniques like Deep Neural Networks and Graph Networks, they are increasingly being used across a myriad of applications including materials discovery [1] medicine [2,3], natural language processing [4], driverless cars [5], and more [6-9]. Given the exponential growth of these AI/ML applications, specialized architectures have been proposed and designed for performance [10, 11, 12, 13]. These specialized architectures aim to optimize computational performance and efficiency for a specific application, which in this case are determined by the AI/ML methods. Our analysis estimates energy for both hardware and algorithms used in systems relevant to AI/ML methods. In this work, we do not address the technology aspects of scaling known as Moore's law, which have been addressed before [14]. The paper is divided into three sections. In the second section, we estimate energy trends associated with computing by analyzing three aspects: energy for instruction execution rates across different architectures, energy from chips to integrated systems, and energy required by large AI/ML applications as applied to Natural Language Processing. In the third section, we analyze the different trends on energy efficiency, driven by the increasing use of AI/ML methods and slowing trends in geometrical scaling. In the final section, we summarize the analysis and highlight the need to consider energy of computing as a key aspect of design.

## II. Energy Trends

Our energy estimate analysis section is divided into three sub-sections covering most of the last decade (2010s through 2022), specifically the period in which AI/ML methods have become widely applied. In the first section below, we analyze the accelerator data of specialized architectures, consisting of specific products, the year of introduction, the transistor count, power, and the instruction execution rates (Instructions per second). In the second section, we analyze supercomputers as another class of specialized systems, taken from the most recent exaflop supercomputer inclusive Top500 list [15]. In the third section, we examine a specific set of accelerators and estimate the energy usage from chip level to rack level. In the last section, we estimate the energies for training AI/ML methods for the specific application of Natural Language Processing (NLP).

### A. Specialized Architectures: AI/ML Accelerators

First, we estimate energy trends in different architectures that are tailored towards AI/ML methods. The systems are designed for high performance (intensity and scale of the number of operations needed), large amounts of data used (related to sensing and edge computing), and the high number of mathematical operations needed (for scientific computations). Although these architectures are specialized, they also include

---

aspects of architecture from general purpose computers [12]. We examine the relevant variables of accelerator systems used in a variety AI/ML edge applications in which both training and inference need significant computations. The GPUs and AI accelerators analyzed include all the major data center accelerator products and are in given in Table 1.

| *Company* | *Accelerator* | *Company* | *Accelerator* |
|---|---|---|---|
| AMD | MI8 | Habana | Gaudi |
| AMD | MI60 | Habana | Goya |
| AMD | MI100 | Huawei | Ascend 910 |
| AMD | MI210 | Intel | Xe-HPC |
| AMD | MI250 | NVIDIA | K20 |
| Baidu | Kunlun K200 | NVIDIA | K80 |
| Cerebras | CS-2 | NVIDIA | P100 |
| Google | TPU1 | NVIDIA | V100 |
| Google | TPU2 | NVIDIA | T4 |
| Google | TPU3 | NVIDIA | A100 |
| Google | TPU4i | NVIDIA | A10 |
| GraphCore | GC2 | NVIDIA | A30 |
| GraphCore | GC200 | Qualcomm | Cloud AI 100 |
| Groq | TSP GroqChip | TensTorrent | Greyskull |

**Table 1**: The list of Accelerators and GPUs analyzed in Figure 1 below.

The energies are estimated per instruction (EPI), and are based on either floating point or integer operations. The energies are computed from the power and the peak number of instructions per second for integer 4-bit (INT4), integer 8-bit (INT8), floating point 16-bit (FP16/BF16), floating point 32-bit (FP32), and floating point 64-bit (FP64). The values are plotted in Figure 1, where the energies are estimated in units of Joules per instruction for each of the execution types listed above. Figure 1 highlights a few important observations. First, scaling in terms of energy efficiency has slowed down compared to ideal geometrical scaling, (assuming two years between succeeding generations, the number of doublings per decade should be approximately close to thirty two). This trend is highlighted by the ratio of maximum to minimum energy/instruction (Figure 1 and the table below the figure), ranging from 6.64 to 26.71 (excluding supercomputers). From Figure 1, it can be seen that the focus and improvements are being increasingly placed on integer instructions, given its relevance to increased AI/ML applications. Finally, EPI follows a sharper reduction than that expected by just technology scaling alone. For example, the ratio of the maximum to minimum EPI over a period of ten years is 6.64 for INT4, while it is 26.71 for FP16 and 13.3 for FP32. This also illustrates the larger reduction in EPI for floating points brought about by combinations of architectures, hardware, and technologies.

As mentioned before, the focus on architecture for the AI/ML applications is increasingly weighted in terms of integer operations resulting in more increases of energy efficiency per instruction compared to that per floating point. This is consistent with the analysis by Hennessy and Patterson [10] that deep neural networks (DNN) use lower precision floating point and integer instructions. Our analysis also indicates that larger energy efficiency improvements are seen across different architectures in a given year than across different technologies over the decade, indicating the shift towards accelerating development of domain-specific specialized architectures.

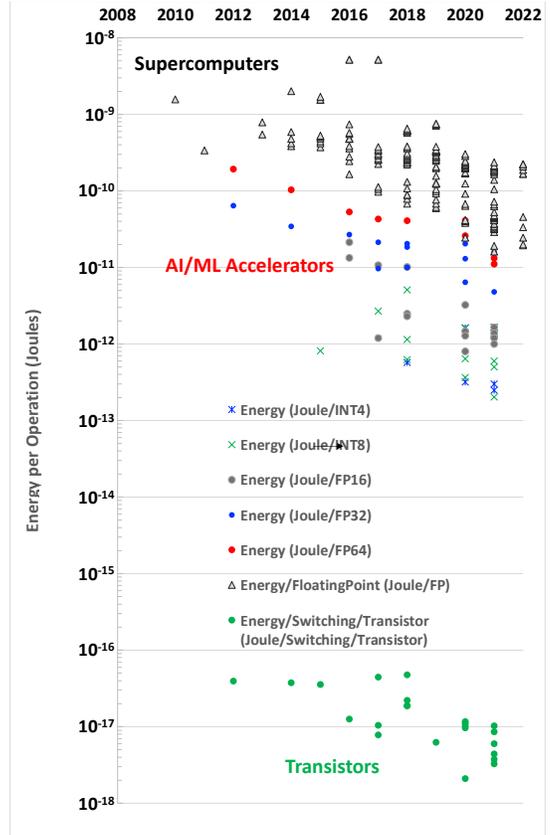

| | Energy (Joule/Switching /Transistor) | Energy (Joule/INT4) | Energy (Joule/INT8) | Energy (Joule/FP16) | Energy (Joule/FP32) | Energy (Joule/FP64) | Energy in SC (Joule/FP) |
|---|---|---|---|---|---|---|---|
| Minimum | 2.1x10^-18 | 2.5x10^-13 | 2.0x10^-13 | 8.0x10^-13 | 4.8x10^-12 | 1.1x10^-11 | 1.6x10^-11 |
| Maximum | 4.7x10^-17 | 1.6x10^-12 | 5.0x10^-12 | 2.1x10^-11 | 6.3x10^-11 | 1.9x10^-10 | 5.1x10^-9 |
| Maximum/Minimum | 22.75 | 6.64 | 24.95 | 26.71 | 13.30 | 17.42 | 321.81 |

**Figure 1**: Energy estimated per operation for different products and processors from 2012 to 2022 in Joules/Instruction. The closed circles represent floating point operations, open symbols represent integer operations per second, and triangles represent the Energy/Floating point (Rmax) for the class of supercomputers [15]. The ratio of maximum to minimum energy estimates for each of the instructions are given in the table below the figure. Higher energy for floating point operations across the years can be observed. Energy trends in Joule/Switching per transistor of specific products in the time frame analyzed are given in green symbols indicate moving downwards to attojoules and discussed in more detail in Section III.B.

*B. Specialized Architectures: Supercomputers*

In this section, we analyze the energy per instruction of the top 500 supercomputers maintained by TOP500 [15]. This list has been compiled twice a year since 1993 and contains the list of the 500 most powerful computer systems with statistics on performance on the high performance LINPACK (HPL) benchmark. Our analysis is based on the 59th edition including what is reported as the first exa-scale computer, but also includes systems introduced between 2010 and 2022 and are still on the list. In this paper, we examined metrics as reported for these classes of machines of the TOP500 HPL (Rmax, Rpeak) and the related high-performance conjugate gradient (HPCG) benchmarks. The HPL benchmark solves a dense linear system of equations in double precision (FP64) arithmetic, using an LU decomposition and back substitution. The central computation is a massive series of matrix multiplications, which can take advantage of many modern CPU, GPU, and memory subsystem innovations for parallel computations and memory accesses. The Rmax is the maximum performance achieved on the HPL benchmark, and Rpeak is the theoretical peak, with values reported in tera FLOPS or peta FLOPS [15,16]. The HPCG benchmark characterizes the data access patterns and computations of a conjugate gradient solver, which underlies many physical system simulations. It involves solving a structured sparse linear system of equations using stencils in double precision (FP64) arithmetic. Because its memory access patterns involve structured sparse accesses, it consistently performs at a lower FLOPS rate than HPL [17]. One point to note is that data are not available for all the metrics across all the five hundred systems. For example, for the exa-scale computer, the HPCG benchmark is not currently available. Also, one should note that both the HPL and HPCG benchmarks require execution in FP64 numerical precision to guarantee an even playing field and numerical stability.

The energy per instruction (EPI) is calculated from the power and the respective measured or estimated values of HPL Rmax, HPL Rpeak and HPCG measured in terms of operations per second. The Energy/Floating Point (Rmax) is plotted in Figure 1 for comparisons between the different AI/ML accelerators and supercomputers. Although there is partial overlap between accelerator and supercomputer classes, the energy per floating point operation is higher for the latter. This could be driven by the complex instructions needed for general purpose scientific computing, along with the fact that the power values for the supercomputers are at system-level while accelerator level power values are taken at the chip- or card-level.

The statistics of all the supercomputers are given in Figure 2, in which the following trends can be seen: the energy per computation in joules/floating point is highest for HPCG, followed by HPL Rmax and system Rpeak. Energy per computation in Joules/floating point range from $1.6 \times 10^{-11}$ to $5.17 \times 10^{-9}$ (Rmax), $1.12 \times 10^{-11}$ to $4.25 \times 10^{-9}$ (Rpeak), and $1.36 \times 10^{-9}$ to $3.20 \times 10^{-8}$ (HPCG). This is because for floating point, Rpeak quantifies peak operation and hence an upper bound for Rmax. The ratio of the number of HPCG operations to Rmax ranges from 13 to 193 for the reported machines indicating higher energies per instruction for HPCG which encompasses lower I/O throughput than that for HPL [17]. Similar data for Rmax and HPCG are plotted for all the individual systems in Figure 3. The estimates indicate that the fastest supercomputer on the leftmost side in the X-axis is relatively energy efficient and is ranked second. It can also be observed that higher energy per operation is required by the HPCG benchmarks compared to HPL.

Among the five hundred supercomputer systems, the maximum EPI for Rmax is $5.17 \times 10^{-9}$, while the minimum EPI is $1.61 \times 10^{-11}$, indicating a range over two orders of magnitude. For the HPCG, this range is over an order of magnitude indicating a slower reduction in energy per instruction. We can compare the supercomputer EPI with that of the accelerators, with the instructions as floating point operations.

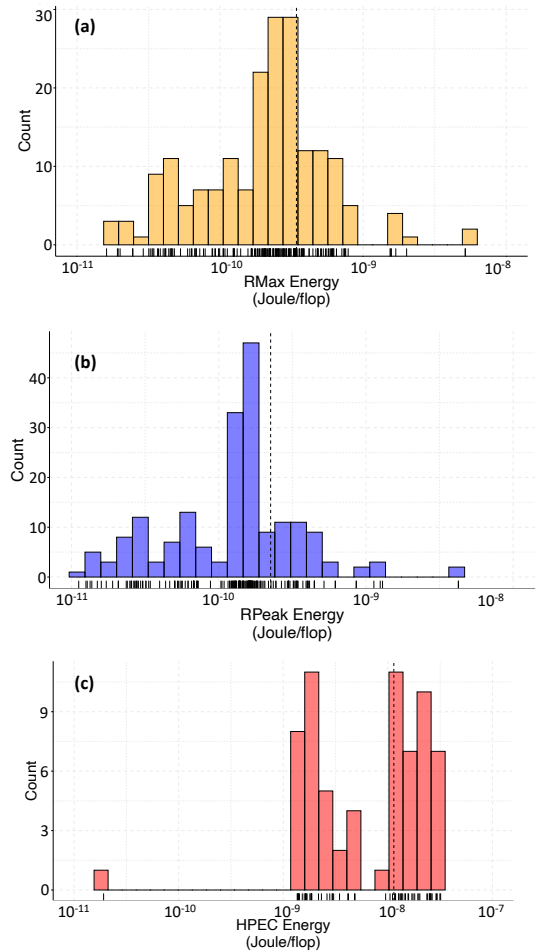

**Figure 2:** Energy/Instruction based on RMax, RPeak, and HPCG for the top 500 supercomputers [15]. The means change for different benchmark metrics.

As seen in Figs. 1-3, the maximum energy (in Joules/floating point) is $1.92 \times 10^{-10}$ (FP64) for accelerators, and $5.17 \times 10^{-9}$ for Rmax in supercomputer systems. There are four orders of magnitude difference in EPI between most energy efficient and the least energy efficient systems across both the accelerator and supercomputers. The trends highlight that typically more energy is needed per operation for high precision computations, as we move from integer to floating point operations or from general purpose benchmark to high performance precision-based computations. The newer systems, including the only exa-scale computer system, tend to be more energy efficient as seen in Figure 4, where we have plotted the Rmax of the first few high end computational systems. Also, while HPL (Rmax) and HPCG are highly correlated (R=0.81), higher energy efficiency in HPL (Rmax) does not necessarily imply corresponding efficiency in HPCG (e.g., compare Lumi or Adastra with Athena in Fig. 4).

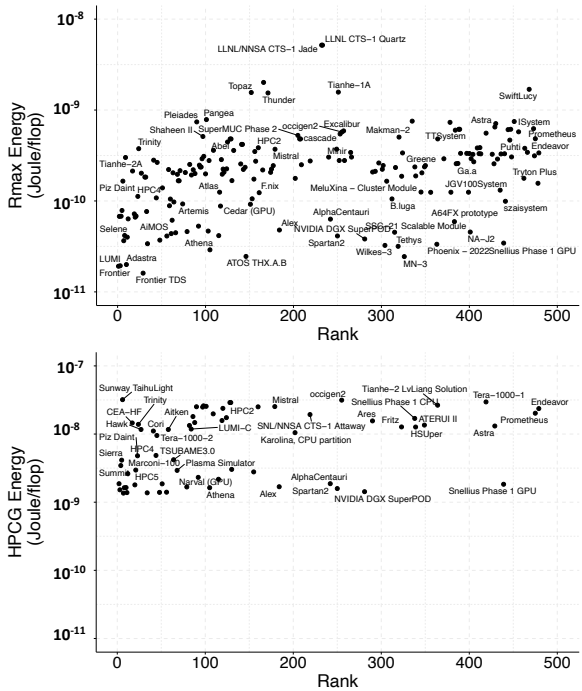

**Figure 3**: Energy/Instruction based on two benchmarks, (a) RMax and (b) HPCG. The X-axis consist of the top 500 supercomputers with the fastest on the left and slowest on the right. Higher energy per instruction for HPCG benchmark is seen in (b) compared to Rmax in (a).

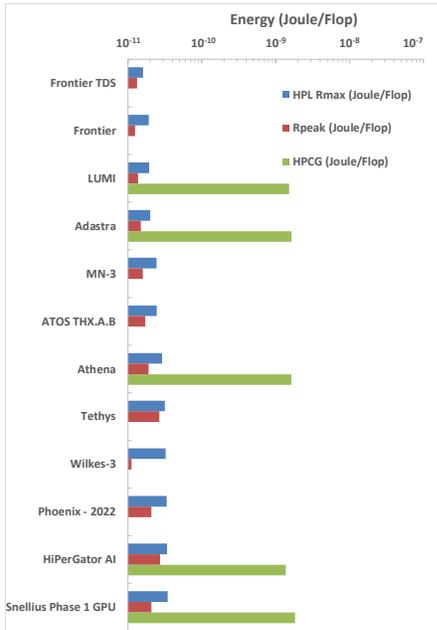

**Figure 4**: Energy/Instruction based on Rpeak, HPL Rmax, and HPEG for the top few supercomputers sorted by energy per instruction based on Rmax. The differences of the benchmarks can be observed (blue-Rmax, red-Rpeak, green-HPCG)

### C. From Chips to Systems

Next, we estimate energy trends in architectures by comparing chips to systems. The data consist of accelerators manufactured across three technologies (from 16 nanometers down to 7 nanometers) and varying integrations of systems (going from chips to racks), with different configurations and architectures [18]. In addition, as memory configuration and the interconnects all determine the energy usage by the system, higher-level analysis alone may not be sufficient to understand the differences between the various design attributes. However, this top-down EPI analysis helps in observing the trends across multiple systems, technologies, hardwares, and designs.

The list of systems analyzed include performance characteristics of chips, servers, and racks and the corresponding metrics based on floating point instructions are plotted in Figure 5. The trends, as shown in Figure 5, indicate that EPI increases from Chip-level to Rack-level, independent of the system variable, technology, hardware, or architecture. The decrease in EPI is an order of magnitude smaller than either that due to technology scaling or across architectures as discussed in the previous section. The key observation is that energy efficiency decreases as more components are integrated on top of the transistors for larger computing systems, underscoring a key challenge in developing energy efficient computing systems from efficient components.

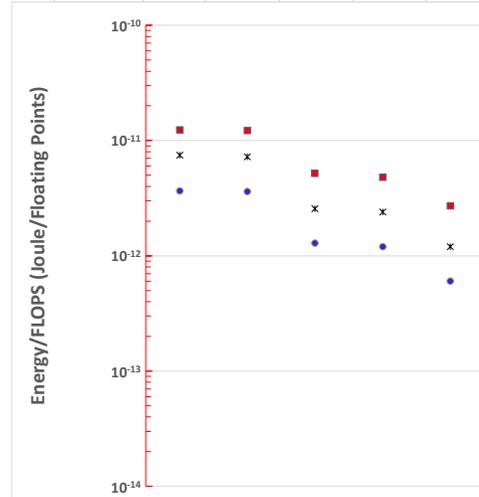

**Figure 5:** Energy (Joule/Floating point) estimated for different systems consisting of Chips, Servers, and Racks. The X-axis represent the different systems as given in Table 2. The gap between Chips and Systems is shown.

### D. Machine Learning in Natural Language Processing

Within AI/ML applications, Nature Language Processing (NLP) is used to parse vast amounts of literature in all languages and also enable computer-aided translation between the different languages. As opposed to linguistical methods, AI/ML methods depend on simpler analysis using words, phrases, part-of speech requirements, existing collections of text, and have been found to be very effective in applying these methods to reproduce texts with reasonable accuracies [4].

Recent analysis indicates that the time it takes for the number of training parameters to double is under eight months since 2016 [19]. In contrast, geometric technology scaling takes more than twenty-four months, resulting in exponential increases of NLP computations ranging from 6 x $10^{18}$ to 3 x $10^{21}$ FLOPs for training a large model with over 100 billion NLP parameters [20, 21]. In Figure 6, we show both the number of floating-point operations needed for training and the number of floating points per parameter from 2012 to 2022. The number of floating-point operations has reached yotta-scale (1 x $10^{24}$) recently. In turn, this increasing number of operations requires proportionately large computational power leading to increased energy consumption per application. This increase in energy consumption is indicative of the floating operations needed for training a single model and does not include inference. The main difference between training and inference is that the former needs large scale, higher precision (FP32 or FP16) computing upfront, while the latter is used routinely in applications many times over, generally at lower precision (FP16, INT16 or INT8). It is conceivable that the computational cost of training will be weighted upfront, while that of inference is spread over the entire time of application with comparable estimates, but further analysis is needed to quantify the differences between the two.

million and 30 billion joules (17 Kilowatt-hour to 8.3 Megawatt-hours). This energy at the lower bound is comparable to the average monthly electricity usage by the city of Atlanta or Los Angeles (~700 Kilowatt-hours in 2017) and at the upper bound compares with the monthly total electricity use by the top fifteen US cities combined (~9 Megawatt-hours in 2017). Clearly, even with the lower estimates, if the current trends were to persist, the energy consumption in training alone will more than off-set the energy efficiency from geometrical scaling or that due to architectures. In fact, more recent analysis suggests that the computational requirements for training NLP models may be many more orders of magnitude higher than the range we have addressed [21-24].

### III. ENERGY EFFICIENCY AND BIT UTILIZATION: CHALLENGES AND OPPORTUNITIES

Examining the trends between different hardware accelerators, supercomputers, systems from chips to racks, and the computational requirements of AI/ML methods in NLP will help in understanding the energy utilization in different aspects of computing needs renewed focus. As a first step, we will analyze power, frequency, and the number of transistors for the different products as illustrated in Figure 7.

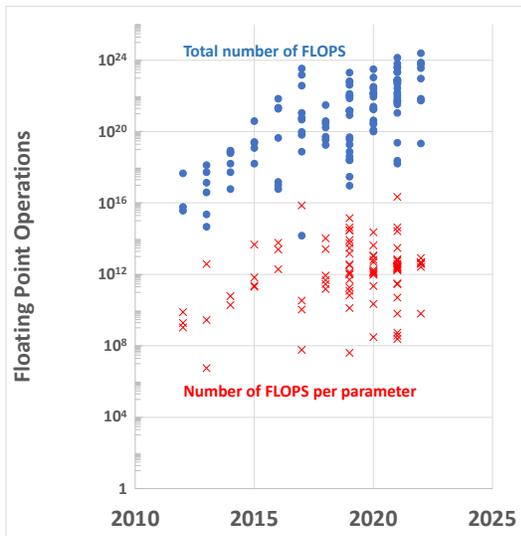

**Figure 6:** Total number of floating point operations (FLOPs) and the number of floating point operations needed per parameter for training NLP models from 2012-2022, based on analysis in [18]. The plot illustrates the exponential increase of computations needed for training in AI/ML applications.

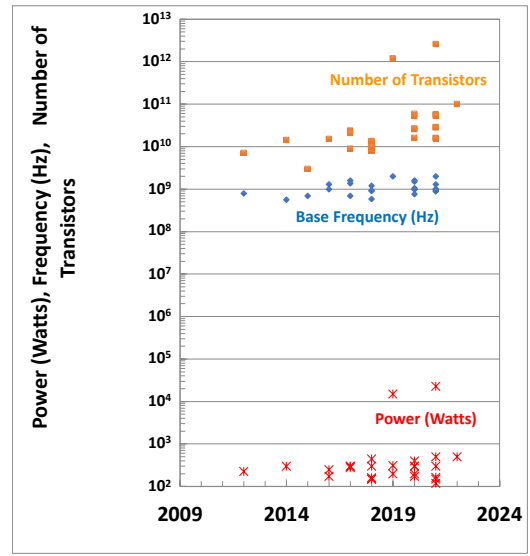

**Figure 7:** Power, frequency and the number of transistors for the different systems considered. Trends show the increasing number of transistors in accordance with Moore's law, while the power and frequency curves are relatively flatter.

Based on the analysis from the previous sections, we can roughly estimate the energy required for computing for a specific application. For illustration, we can assign EPI of 1 x $10^{-12}$ Joule/FPI (FP16) (lower bound) to 1 x $10^{-11}$ (FP64) (upper bound) representing two energy efficient floating-point cases from Figure 1. If the number of floating-point operations needed for training of a single model of NLP is between 6 x $10^{18}$ and 3 x $10^{21}$, the energy required for training would be between 6 million and 3 billion joules (1.7 Kilowatt-hour to 833 Kilowatt-hours) for lower bound, while the estimates are for 60

Over the period that we have analyzed, number of transistors has been increasing slower than that resulting from a two-year cadence of geometrical scaling (~20X compared to ~32X/decade without including the two very large microprocessors). If we include larger multi-die compute-intensive chips like Cerebras CS-2, AMD MI250, and Intel XE HPC, which have integrated over 100 billion transistors, they exceed the conventional trends of doubling of the number of transistors every few years. Unlike the number of transistors, power and frequency increases are relatively flatter indicating energy management for an increased transistor budget (~3X to 6X). This indicates a general slowing down of the traditional geometrical scaling every two years. In the next section, we will analyze the relationships between bits and instructions and use it to formulate pointers for increasing energy efficiencies beyond geometrical scaling.

*A. BIT Utilization*

A bit switching refers to transistor operation switching from high to low voltage or vice versa or between any two distinct states for classical computing. The number of bits switching per second relates to the frequency at which the chip is being clocked and subsequently the switching rate of the transistors determining elementary information processing rates. At the system level, the corresponding variable is the number of instructions per second, which is enabled by multiple elementary information processing rates. Since the instructions can be expressed in terms of integers or floating point operations, there is a corresponding metric for each of the instructions: INT'X' is the number of X-bit integer instructions per second (where X is 4 or 8 bits), while FP'Y' is Y-bit floating point instructions per second (where Y is 16, 32, or 64 bits). Comparing the transistor-level switching to instructions per second can help in estimating how the bits and transistors are utilized at the system level. These relationships have been also applied elsewhere to analyze trends in system level performance [25]. For example, if the instructions per second (IPS) for the system are the same as the number of bits switching of all the transistors (BPS), then the bit utilization (BPS/IPS) will be unity. In other words, all the bits at elementary information processing are proportionally utilized at the system level instructions for computing. In Figure 8, we plot the IPS (Y-axis) versus BPS (X-axis) for the same accelerator products from Table 1 and in Figure 9, the Bits/Instructions are plotted versus BPS.

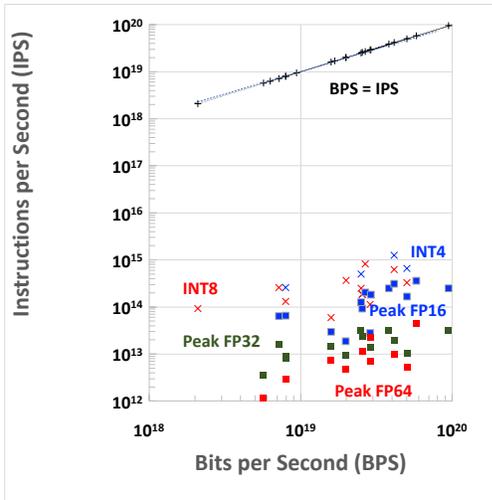

**Figure 8:** Instructions per second for both integer and floating points operations (IPS) are plotted against bits per second in the x-axis (BPS) for accelerator products from Table 1. IPS = BPS curve is marked in black symbols. The plot highlights the gap between bit switching and instruction executions.

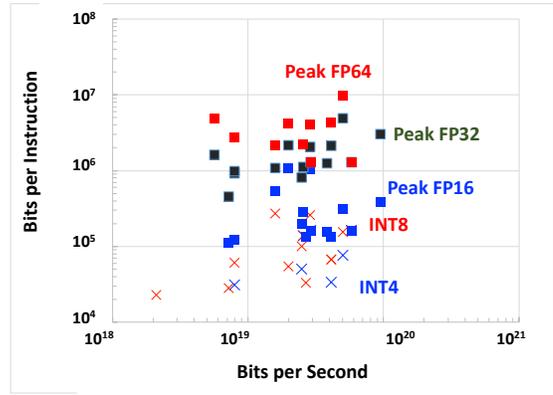

**Figure 9:** Bits per Instruction (BPI) for different instructions are plotted against Bits per second (BPS) in the x-axis for accelerator products from Table 1. More bits (and transistors) are needed for floating point instructions (FP16, FP32, FP64) than for integer instructions (INT4 and INT8).

A few key observations from this analysis can be summarized as below. As the bit utilization increases, the points are closer to this curve, as seen for INT4 and INT8, which are increasing with the current focus on new accelerator architectures. The floating point-based instructions (FP32 and FP64) are farther from the unity curve indicating less utilization of the bits at the system level which needs more bits and transistors for each of the instructions. This makes intuitive sense because floating point operations are more complex and require more complex circuits of transistors compared to integer operations. The disparity between EPI at the system level and energy per bit at the transistor level is large, indicating that more bits are needed to compute applications, as analyzed in the next section. In Figure 9, the numbers of bits and transistors needed for an instruction at the system-level highlights two points: 1) The number of bits per instructions are higher for floating points than integer operations; 2) As the number of bits per second (BPS) increases due to either higher frequencies or more transistors, the bits per instructions do not generally decrease. The second point indicates the increasing complexity at the architecture level or the application (AI/ML in this case) needs more bits and transistors. This could be attributed to a variety of factors such as: storing and accessing data from high-speed memories including on-chip Static Random Access Memory (SRAM), itself needing six transistors to read, write, and store; interconnects which ferry bits back and forth in addition to information processing units; off-chip Dynamic Random Access Memory (DRAM), etc. As the applications need complex operations or higher precision for computations, the energy usage increases are driven by intrinsic factors of Boolean logic-based information processing.

*B. Energy Efficiency Trends*

The energy per switching per transistor, which is energy per bit (EPB) in joules, is given in Figure 1 (as green symbols), observed to be lower by ~10X over the time of our analysis. This result serves as a metric of transistor-level energy efficiency gained over the decade. Comparing the energy for switching at the transistor level to that at the instruction level (Figure 1) reveals the following observations: Energy per instruction (EPI) is higher than EPB by 2.3 x $10^4$ to 9.7 x $10^6$. It is also clear that integer operations are more energy efficient than floating point operations per instruction, which is consistent with increasing efforts on optimizing integer instructions in accelerator architectures and the above observations regarding bit utilization.

As indicated above, the energy efficiency decreases when transistors are integrated with more components into a system for computations. In addition, the innovations in architectures and the relatively lower precision requirements like integer or lower precision floating point operations such as FP16 seem to be driving the architecture changes. However, this computational efficiency is offset by the increase in complexity of modeling algorithms as exemplified by the one trillion parameters for machine learning in natural language processing. During training, the number of floating-point operations per parameter can be higher than 1 x $10^{12}$; i.e., a trillion floating point operations are needed to characterize a single floating-point parameter. This shows that despite innovations within the computing ecosystem, the AI/ML models at least for Nature Language Processing are computationally expensive in addition to being energetically inefficient in many cases. Moreover, these results, such as those shown in Figure 1, suggest future innovations in energy efficient computing require new framing of designing algorithms and software for Machine Learning for every application rather than general-purpose architectures with general purpose AI/ML algorithms, consistent with other recent analysis [26]. In addition to the new transistors, systems, and technology scaling, developing specialized algorithms and software that are also energy efficient will require integrating different perspectives. This will enable a sustainable path in the future of computing, further reinforcing the need for systematic Co-design across all aspects [14].

Additionally, our analysis of NLP models hints at the difficulty in applying current AI/ML architectures for scientific problems. Although the corpus used in language models are large as in 170,000 words in the English language, the complexities of scientific problems are even much larger. For example, the number of possible organic chemicals can be higher than $10^{30}$ [27]. Using the same ratio of training floating point operations in the previous example of a trillion operations per parameter, the number of floating-point operations needed for training AI/ML models for chemical systems can exceed by several orders of magnitude over that of NLP models. As a result, the corresponding energy requirements could exceed thousands of kWh, which is more than the total monthly electricity used by the largest cities in the US. Even without including inference, the higher computational intensity and energy requirements along the current trajectories are amplified when machine learning is extended to multiple scientific disciplines. Because of these trends, incorporating considerations of energy efficiency will be critical for achieving wider usage of AI/ML both from practical and sustainable perspectives.

## IV. CONCLUSIONS

We have estimated the energy requirements of computing in different systems from chips to racks, with a focus on 2012 and beyond when machine learning and artificial intelligence started leading the field of computing applications. Our analysis of systems based on accelerators in AI/ML applications and supercomputers highlights three key aspects: One, energy efficiency due to geometrical scaling is slowing down as has been observed earlier due to increasing complexity in design, processing, integration, and manufacturing [14, 28]. Second, the energy efficiency at the bit-level does not map directly to the instructions at the system level, in spite of more transistors being integrated in every succeeding generation of technology scaling with minimal change in total power. This can be attributed to the lower bit utilization per instruction (across integer, floating point, and complex benchmarks for scientific computing), which results from the architecture and the communication between computing and memory/storage. As a result, the factors that determine energy per instruction involves more than efficient transistor switching. In addition, the innovations in architectures in a given year can provide higher energy efficiency than that obtained by geometrical scaling alone during the decade that we have analyzed. Third, AI/ML methods for advanced applications like natural language processing and digital automation, have become one of the biggest drivers of computing using artificial intelligence-based methods into all aspects of modern economy. However, these general-purpose AI/ML methods can be computationally energy intensive, off-setting the energy efficiencies achieved by geometrical scaling and/or by design of special purpose accelerators, or integration of specialized large-scale architectures.

The compounding effects of the above factors, catalyzed by the digitization of the economy and ubiquitous availability of data analytic tools, computing devices, and sensors, have led to an increase in energy requirements of computing systems. However, many of the real problems in sciences and engineering are significantly higher in terms of complexity than language processing models. This higher complexity indicates that development and training of AI/ML models for scientific and engineering problems are likely to be very energy intensive as well. As a result, we believe that in order to attain sustainable computing with practical constraints, it is important to use energy as an additional design variable that bridges architecture and algorithms in addition to hardware and technology. This may need a different framing of computing from a general-purpose solution to bespoke computing systems that are finely tuned end-to-end, specialized for every application using Co-design, with energy efficiency as a key attribute.

ACKNOWLEDGMENT

We appreciate the information provided by M. Khairy and T. Rogers (Purdue), Tamay Besiroglu (MIT) for clarifications and access to their information, and V. Zhirnov (SRC) for discussions on bit utilization. SS also appreciates the institutional support from SLAC National Laboratory (C.C. Kao, P. McIntyre) and MIT LLSC (J. Kepner).

This work was partially supported by the U.S. Department of Energy's Office of Science contract DE-AC02-76SF00515 with SLAC through an Annual Operating Plan agreement WBS 2.1.0.86 from the Office of Energy Efficiency and Renewable Energy's Advanced Manufacturing Office. The Accelerator data analysis is based upon work supported by the Assistant Secretary of Defense for Research and Engineering under Air Force Contract No. FA8702-15-D-0001. Any opinions, findings, conclusions or recommendations expressed in this material are those of the author(s) and do not necessarily reflect the views of the Assistant Secretary of Defense for Research and Engineering.